\documentclass[prb,superscriptaddress,preprint,endfloats,nopacs]{revtex4}

\usepackage{amsmath, amsthm, amssymb, pifont, wasysym}

\usepackage{comment}
\usepackage{graphicx}
\usepackage{bm} 
\usepackage{pifont}
\usepackage{dcolumn} 
\usepackage{color}
\usepackage{siunitx}
\usepackage{float}
\definecolor{mygray}{gray}{0.5}

\newcommand {\CSS}{Co$_3$Sn$_2$S$_2$}
\newcommand {\CMG}{Co$_2$MnGa}
\newcommand {\MS}{Mn$_3$Sn}
\newcommand {\ECS}{EuCd$_2$Sb$_2$}
\newcommand {\ECA}{EuCd$_2$As$_2$}
\newcommand {\ETO}{EuTiO$_3$}

\newcommand {\TN}{$T_{\mathrm{N}}$}

\newcommand{\sigmaxx}{$\sigma_{xx}$}
\newcommand{\sigmaxxB}{$\sigma_{xx}^{\bm{B}(1)}$}
\newcommand{\sigmaxxBsecond}{$\sigma_{xx}^{\bm{B}(2)}$}
\newcommand{\sigmaxyAHE}{$\sigma_{xy}^{\rm{AHE}}$}

\newcommand{\dsigmaxyAHE}{$\varDelta$$\sigma_{xy}^{\rm{AHE}}$}
\newcommand{\magnetoresistivityratio}{$(\rho_{xx}-\rho_{xx,\rm{0 T}})/\rho_{xx,\rm{0 T}}$}

\newcommand{\rhoyxAHE}{$\rho_{yx}^{\rm{AHE}}$}

\begin{document}
	
	\title{Berry curvature derived negative magnetoconductivity \\
		observed in type-II magnetic Weyl semimetal films}
	
	\author{Ayano Nakamura}
	\affiliation{Department of Physics, Tokyo Institute of Technology, Tokyo 152-8551, Japan}
	\author{Shinichi Nishihaya}
	\affiliation{Department of Physics, Tokyo Institute of Technology, Tokyo 152-8551, Japan}
	\author{Hiroaki Ishizuka}
	\affiliation{Department of Physics, Tokyo Institute of Technology, Tokyo 152-8551, Japan}
	\author{Markus Kriener}
	\affiliation{RIKEN Center for Emergent Matter Science (CEMS), Wako 351-0198, Japan}
	\author{Mizuki Ohno}
	\affiliation{Department of Applied Physics, The University of Tokyo, Tokyo 113-8656, Japan}
	\author{Yuto Watanabe}
	\affiliation{Department of Physics, Tokyo Institute of Technology, Tokyo 152-8551, Japan}
	\author{Masashi Kawasaki}
	\affiliation{RIKEN Center for Emergent Matter Science (CEMS), Wako 351-0198, Japan}
	\affiliation{Department of Applied Physics, The University of Tokyo, Tokyo 113-8656, Japan}
	\author{Masaki Uchida}
	\email[Author to whom correspondence should be addressed: ]{m.uchida@phys.titech.ac.jp}
	\affiliation{Department of Physics, Tokyo Institute of Technology, Tokyo 152-8551, Japan}

	\begin{abstract}
	Here we study nonmonotonic features which appear both in magnetoresistivity and anomalous Hall resistivity during the simple magnetization process, by systematically measuring type-II magnetic Weyl semimetal {\ECS} films over a wide carrier density range. We find that a positive magnetoresistivity hump can be explained as manifestation of a field-linear term in the generalized magnetoconductivity formula including the Berry curvature.  As also confirmed by model calculation, the term can be negative and pronounced near the Weyl point energy in the case that the Weyl cones are heavily tilted. Our findings demonstrate extensive effects of the Berry curvature on various magnetotransport in magnetic Weyl semimetals beyond the anomalous Hall effect.
	\end{abstract}
	
	\maketitle
	

	Weyl semimetals (WSMs) host pairs of three-dimensional band crossing points called Weyl points in their bulk band structure, where the spin degeneracy is resolved by breaking of either space-inversion or time-reversal symmetry \cite{Weyl_proposed_theory, Weyl_spin_degeneracy, Weyl_linear_dispersion}.
	Magnetotransport reflecting the Berry curvature has recently been attracting great interest in condensed matter physics  \cite{Boltzmann_equation_1}. Owing to the large Berry curvature distribution around the Weyl points, magnetic WSMs with time-reversal symmetry breaking exhibit giant magnetoelectric responses such as in anomalous Hall effect (AHE) \cite{Weyl_AHE_1, Weyl_AHE_2, Weyl_AHE_3}, anomalous Nernst effect (ANE) \cite{Mn3Sn_ANE,ANE_Co2MnGa}, chiral anomaly \cite{chiral_anomaly_1, chiral_anomaly_2, chiral_anomaly_3, chiral_anomaly_4}, and magneto-optical effect \cite{GMO_Co3Sn2S2}. It has been also theoretically proposed that nontrivial coupling between charge and magnetization originating in the chiral anomaly enables further versatile functionalities such as charge-induced spin torque and electric field-induced magnetization switching \cite{coupling_magnetization_and_charge, coupling_magnetization_and_charge_2}.
	
	For magnetic WSMs, 3$d$-electron systems including {\CSS} \cite{Co3Sn2S2_1, Co3Sn2S2_2, Co3Sn2S2_3, Co3Sn2S2_4, Co3Sn2S2_5, Co3Sn2S2_6, Co3Sn2S2_7}, {\CMG} \cite{Co2MnGa_1, Co2MnGa_2, Co2MnGa_3, Co2MnGa_4, Co2MnGa_5,Co2MnGa_6}, and {\MS} \cite{Mn3Sn_1, Mn3Sn_2, Mn3Sn_3, Mn3Sn_4, Mn3Sn_5} have been intensively studied so far. 
	AHE and also ANE are largely enhanced due to the presence of multiple Weyl points in the ferromagnetic ground state and observed up to high temperatures near or well above room temperature \cite{Weyl_AHE_2, Mn3Sn_ANE, ANE_Co2MnGa, Co2MnGa_film}. However, their band structures with many Weyl and non-Weyl bands near the Fermi level are too complex to accurately evaluate magnetotransport derived from the Weyl points. On the other hand, 4$f$-electron magnetic WSMs including GdPtBi \cite{GdPtBi_1, GdPtBi_2, GdPtBi_3, GdPtBi_4}, {\ECS} \cite{EuCd2Sb2_crystal_structure, EuCd2Sb2_1, EuCd2Sb2_2, EuCd2Sb2_3, EuCd2Sb2_4}, and possibly {\ECA} \cite{EuCd2Sb2_crystal_structure, EuCd2As2_1, EuCd2As2_2, EuCd2As2_3} exhibit a more simple band structure, where only the bands forming the Weyl points exist near the Fermi level in the forced ferromagnetic state. Moreover, their semimetallic nature with low carrier densities (\num{e18}--\num{e19} \si{cm^{-3}}) as compared to those of 3$d$-electron systems (\num{e20}--\num{e22} \si{cm^{-3}}) make it possible to elucidate the carrier density dependence across the Weyl point energy. Recently, the detailed Fermi level dependence of AHE has been examined by measuring {\ECS} thin films with incorporating electrostatic gating \cite{EuCd2Sb2_film_1}.
	
	{\ECS} may be a suitable system for studying Berry curvature derived magnetotransport in type-II magnetic WSMs \cite{typeII_WSM}. It has a trigonal crystal structure with space group $P\bar{3}m1$, as sketched in Fig. 1(a) \cite{EuCd2Sb2_crystal_structure}. The Cd 5$s$ and Sb 5$p$ bands are inverted to form a simple band structure with type-II band crossings near the Fermi level, and the Eu 4$f$ states provide large isotropic magnetic moments (Eu$^{2+}$: $S=7/2, L=0$) \cite{EuCd2Sb2_1, EuCd2Sb2_2}. At zero magnetic field, in-plane A-type antiferromagnetic ordering is realized below the N\'{e}el temperature {\TN} = 7.5 K, where a gap opened by the antiferromagnetic ordering is negligibly small and the crossing points can be seen almost degenerate as Dirac points \cite{EuCd2Sb2_2}. By applying the magnetic field $\bm{B}$ along the $c$-axis, the magnetic moments are gradually canted without spin-flop transition and are fully aligned along the $c$-axis above the saturation field $B_\mathrm{s}$ \cite{EuCd2Sb2_1}. During the simple magnetization process, a few pairs of Weyl points are formed by splitting of the Dirac points and shifted in energy and momentum with the Zeeman band splitting.
	
	While magnetotransport of 4$f$-electron magnetic WSMs has been studied mainly focusing on the forced ferromagnetic state, intriguing features may appear also during the magnetization process. One typical example is the nonmonotonic behavior observed in the anomalous Hall resistivity below $B_\mathrm{s}$ \cite{EuCd2Sb2_film_1, nonlinear_AHE_EuCd2As2, AHE_EuTiO3, GdPtBi_peak}. Furthermore, it has been theoretically proposed that the Berry curvature can also affect magnetoconductivity or magnetoresistivity especially when the Weyl cones are tilted as in type-II WSMs \cite{current_berry_curvature}.
	
	Here we systematically study magnetotransport of {\ECS} thin films over a wide carrier density range, particularly focusing on nonmonotonic features appearing below the saturation field. In addition to the nonmonotonic feature in the anomalous Hall resistivity, we find that a broad hump appears in the magnetoresistivity curve. This can be interpreted as manifestation of a field-linear term in the generalized magnetoconductivity formula including the Berry curvature.

	{\ECS} single-crystalline films were grown on CdTe (111)A substrate using a molecular beam epitaxy (MBE) technique \cite{EuCd2Sb2_film_1}. In order to remove native oxide and obtain an atomically smooth surface, CdTe substrates were etched with 0.01 \% Br$_2$-methanol and heated to 680 \si{\degreeCelsius} under Cd flux inside the MBE chamber prior to the growth \cite{CdTe_etching, Sb_doped_Cd3As2}. {\ECS} films were then grown at 324\si{\degreeCelsius} by codeposition of Eu, Cd, and Sb with Cd rich pressure ratios ($P_{\mathrm{Cd}}$/($P_{\mathrm{Eu}}$+$P_{\mathrm{Sb}}$) $\sim$ 30 - 60). For achieving high crystallinity, the films were annealed at 453 \si{\degreeCelsius} for 3 minutes with supplying excess Cd after the growth. Growth orientation is along the $c$-axis, and the film thickness was designed at 50 nm. 
	 Since Cd is easily deficient in {\ECS}, we have obtained {\ECS} films with hole densities ranging from \num{e18} to \num{e20} \si{cm^{-3}}.
	 Longitudinal resistivity {$\rho_{xx}$} and Hall resistivity {$\rho_{yx}$} were measured with employing a standard four-probe method on a Hall bar. Aluminum wire was connected to the terminals by using an ultrasonic bonding machine and these connections were reinforced by appending silver paste. Low-temperature measurements of {$\rho_{xx}$} and {$\rho_{yx}$} up to 9 T were performed using a Cryomagnetics cryostat system equipped with a superconducting magnet. 
	 Magnetization curves up to 7 T were measured using a superconducting quantum interference device magnetometer in a Quantum Design Magnetic Property Measurement System (MPMS).
	 Theoretical analysis on the transport properties were conducted numerically using on-premise pc clusters.
	
	Figure 1(b) presents magnetization curves taken for a {\ECS} film with a hole density of $p$$_{\rm{3D}}$ = \num{0.53e19} \si{cm^{-3}}, measured with applying the magnetic field along the $c$-axis at various temperatures. At the base temperature of $2$ K, the forced ferromagnetic state is realized above the saturation field $B$$_{\rm{s}}$ = $3.5$ T. Below $B_\mathrm{s}$, out-of-plane magnetization increases linearly upon increasing the field, confirming that there are no other magnetic transitions also in the thin film form. Figure 1(c) shows temperature dependence of resistivity taken for three {\ECS} films with different hole densities of $p$$_{\rm{3D}}$ = 0.53, 5.0, and \num{11e19} \si{cm^{-3}}. All the films exhibit a kink structure at $7.5$ K independent of the carrier density, consistent with {\TN} previously reported for bulk single crystals \cite{EuCd2Sb2_1, EuCd2Sb2_2}.
	
	Figure 2 summarizes the magnetoresistivity ratio {\magnetoresistivityratio} and the anomalous Hall resistivity $\rho_{yx}^{\rm{AHE}}$ taken for these three {\ECS} films under the out-of-plane field at various temperatures. $\rho_{yx}^{\rm{AHE}}$ is obtained by subtracting the ordinary Hall term from the raw $\rho_{yx}$ curve. No AHE appears at zero field, since the Berry curvature is zero in total in the antiferromagnetic ground state. When the magnetic field is applied along the $c$-axis direction, $\rho_{yx}^{\rm{AHE}}$ nonmonotonically increases and then decreases with a hump structure, and eventually saturates above $B_\mathrm{s}$. Detailed behavior is dependent on the hole density, while, again, no other magnetic transitions are confirmed. Therefore, this nonmonotonic dependence is interpreted as a change in the electronic structure and Berry curvature during the magnetization process, where the  Weyl points are formed and shifted across the Fermi level with the Zeeman band splitting. This nonmonotonic dependence of {\rhoyxAHE} with the energy shift of the Weyl points has been also theoretically demonstrated for antiferromagnetic {\ETO}, in which the Weyl points are similarly formed and shifted in the Zeeman splitting process \cite{AHE_EuTiO3}.
	
	As confirmed in the upper panel of Fig. 2(a), nonmonotonic behavior or a broad hump structure emerges also in the magnetoresistivity below $B$$_{\rm{s}}$. The field where the hump appears in {\magnetoresistivityratio} is slightly lower than that at which the hump appears in $\rho_{yx}^{\rm{AHE}}$. On the other hand, both the humps similarly shift to lower fields upon increasing temperature and almost disappear above {\TN} = 7.5 K. As seen in Figs. 2(b) and 2(c), the same trend is commonly observed also in the other two films with higher hole densities. This suggests that the hump in {\magnetoresistivityratio} also reflects a field-induced change in the Berry curvature. \color{black} Strictly, in some samples such as the film with $p_{\mathrm{3D}} = 5.0 \times 10^{19}$ $\mathrm{cm}^{-3}$, the hump structure in $\rho_{yx}^{\rm{AHE}}$ remains visible up to about 10 K. This also suggests that the nonmonotonic feature is derived from the momentum-space band Berry curvature, because in this case the finite change can be induced by the magnetic field even above {\TN}. Another hump structure only seen in the magnetoresistivity near zero field is probably ascribed to a zero-field centered negative cusp structure caused by weak anti-localization effect \cite{WAL_1, WAL_2}. \color{black}

	Figure 3 shows hole-density dependence of {\magnetoresistivityratio}, {\rhoyxAHE}, converted magnetoconductivity $\sigma_{xx}$, and nonmonotonic anomalous Hall conductivity {\dsigmaxyAHE}. As shown in Figs. 3(a) and 3(b), both {\magnetoresistivityratio} and $\rho_{yx}^{\rm{AHE}}$ humps are systematically observed in all the films and gradually shift to lower fields with increase in the hole density. These trends also suggest that both of them originate in the change in the Berry curvature. When converted to {\sigmaxx} and anomalous Hall conductivity {\sigmaxyAHE}, {\sigmaxx} shows a negative hump at the original hump position in $\rho_{xx}$, while {\sigmaxyAHE} still shows a positive hump. \color{black} For the high hole density samples, a negative hump also appears at lower fields in the nonmonotonic component of the anomalous Hall conductivity {\dsigmaxyAHE}, as shown in Fig. 3(d). This also indicates that the nonmonotonic feature is derived from the momentum-space band Berry curvature, not from the spin Berry phase reflecting the real-space magnetic ordering. \color{black} The nonmonotonic feature of {\sigmaxx} can be confirmed also in previous reports of other type-II magnetic WSMs \cite{EuCd2As2_peak, GdPtBi_peak}. However, there had been little discussion about its origins.

	To discuss the Berry curvature effect on the magnetoconductivity, here we refer to a generalized formulation derived by solving the Boltzman equation with incorporating the Berry curvature \cite{Boltzmann_equation_1}. According to this, the current density $\bm{J}^{\bm{B}(1)}$ and $\bm{J}^{\bm{B}(2)}$, proportional to the first \cite{current_berry_curvature} and second \cite{current_B_second} powers of the magnetic field, are expressed as 	
	\begin{equation}
		\begin{split}
			\bm{J}^{\bm{B}(1)}&=e^{3}\tau\sum_{n=\pm}\int\frac{dp^{3}}{(2\pi)^{3}}\bm{W}_{\bm{p}n}(\bm{E}\cdot\bm{v}_{\bm{p}n})(f_{\bm{p}n}^{0})'\\
			&-e^{3}\tau\sum_{n=\pm}\int\frac{dp^{3}}{(2\pi)^{3}}(\bm{B}\cdot\bm{E})(\bm{b}_{\bm{p}n}\cdot\bm{v}_{\bm{p}n})\bm{v}_{\bm{p}n}(f_{\bm{p}n}^{0})',
		\end{split}
	\end{equation}
	\begin{equation}
		\bm{J}^{\bm{B}(2)}=-e^{4}\tau\sum_{n}\int\frac{dp^{3}}{(2\pi)^{3}}[\bm{W}_{\bm{p}n}(\bm{E}\cdot\bm{W}_{\bm{p}n})](f_{\bm{p}n}^{0})',
	\end{equation}
	where $(f_{\bm{p}n}^{0})'$ is the energy derivative of the Fermi-Dirac distribution function at zero temperature with momentum $\bm{p}$ and band index $n$, and  $\bm{W}_{\bm{p}n}$ represents $\bm{b}_{\bm{p}n}\times(\bm{v}_{\bm{p}n}\times\bm{B})$ with the Berry curvature $\bm{b}_{\bm{p}n}$ and the velocity $\bm{v}_{\bm{p}n}$. 
	The term $\bm{J}^{\bm{B}(1)}$ is finite only when the time-reversal symmetry is broken and the velocity is anisotropic with tilting of the Weyl cones, and so it is expected to be a leading term in the case of type-II WSMs \cite{current_berry_curvature}. The $\bm{J}^{\bm{B}(2)}$ term is always finite, giving general formula of the chiral anomaly. For the present case of $\bm{E} = (E_{x}, 0, 0)$ and $\bm{B} = (0, 0, B_{z})$, the magnetoconductivity {\sigmaxxB} and {\sigmaxxBsecond} are simply obtained as 
	\begin{equation}
	\sigma_{xx}^{\bm{B}(1)}=e^{3}\tau\sum_{n}\int\frac{dp^{3}}{(2\pi)^{3}}v_{x}^2b_{z}B_{z}(f_{\bm{p}n}^{0})',
\end{equation}
\begin{equation}
	\sigma_{xx}^{\bm{B}(2)}=-e^{4}\tau\sum_{n}\int\frac{dp^{3}}{(2\pi)^{3}}v_{x}^2b_{z}^2B_{z}^2(f_{\bm{p}n}^{0})'.
\end{equation}
 Importantly, {\sigmaxxBsecond} is always positive, but {\sigmaxxB} can be positive or negative, depending on the sign of $b_{z}$ relative to $B_{z}$, namely the splitting direction of the paired Weyl points and also the tilting direction of the Weyl cones. Since the sign of $b_{z}$ changes upon reversal of $B_{z}$, {\sigmaxxB} is also even for the reversal of $B_{z}$. Also note that the anomalous Hall conductivity $\sigma_{xy}^{\rm{AHE}}$ can be expressed as 	
	\begin{equation}
		\sigma_{xy}^{\rm{AHE}}=-e^{2}\sum_{n}\int\frac{dp^{3}}{(2\pi)^{3}\hbar^{4}}b_{z}f_{\bm{p}n}^{0}.	
	\end{equation}
In both the expressions of $\sigma_{xx}^{\bm{B}(1)}$ and $\sigma_{xy}^{\rm{AHE}}$, $b_{z}$ contributes as a linear effect. While $b_{z}$ is integrated only at $E_{\rm{F}}$ in the equation of {\sigmaxxB}, it is integrated fully below $E_{\rm{F}}$ in {\sigmaxyAHE}. 
	
	We further examine the relation between {\sigmaxxB} and {\sigmaxyAHE} by calculating their Fermi energy dependence. We consider a dipolar model for type-II WSM, whose Hamiltonian reads
	\begin{equation}
H=-\frac{k^2}{2m}+ v k_{x} \sigma_{x} + v k_{y} \sigma_{y} + \frac{v}{2}(k_{z}^{2}-k_{0}^{2}) \sigma_{z}.
	\end{equation}
Figure 4 shows the results for magnetoconductivity and anomalous Hall conductivity. They are nonzero below the chemical potential of $\mu \sim 0.14$ above which the hole bands are fully filled. {\sigmaxxB} exhibits a sharp negative peak almost at the Weyl point energy. On the other hand, {\sigmaxyAHE} shows a broad positive hump, which takes its maximum slightly above that energy due to rather complex distribution of the Berry curvature in the type-II WSM. {\sigmaxxB} appears nearly as the energy derivative of $-\sigma_{xy}^{\rm{AHE}}$, which is also easily confirmed by comparing Eqs. (3) and (5). We can confirm qualitative agreement with the experimental results including the signs of {\sigmaxxB} and {\sigmaxyAHE} near the Weyl points energy. Although it is difficult to directly compare to the nonmonotonic features observed in the field sweep, the calculated energy dependence can be roughly viewed as field dependence, where the pair of Weyl points crosses the Fermi level from top to bottom with the Zeeman band splitting and the negative hump in $\sigma_{xx}$ and the positive hump in {\sigmaxyAHE} successively appear. \color{black} Considering the first-principles calculations in {\ECS} bulk \cite{EuCd2Sb2_2} and thin film \cite{EuCd2Sb2_film_1}, we can roughly estimate that the energy shift during the magnetization process from the antiferromagnetic state to the ferromagnetic state by the Zeeman splitting is about 5--10 meV. It is large enough to consider that the Fermi level crosses the Weyl points during the magnetization, leading to a large nonmonotonic change in the Berry curvature. \color{black} Importantly, {\ECS} has small Fermi surfaces with large contribution from the Weyl points, and the Weyl cones are heavily tilted with highly anisotropic velocity \cite{EuCd2Sb2_1}. Therefore, nonmonotonic changes in {\sigmaxxB} and {\sigmaxyAHE} are prominently observed in {\ECS} thin films.
	
	In summary, we have investigated the nonmonotonic behavior which appears both in magnetoresistivity and anomalous Hall resistivity below the saturation field, by systematically measuring type-II magnetic WSM {\ECS} thin films over a wide carrier density range. 
	The positive magnetoresistivity hump can be interpreted as manifestation of the negative {\sigmaxxB} term in the generalized magnetoconductivity formula including the Berry curvature. This term becomes prominent when the Weyl cones are tilted, and also it can be distinguished from the {\sigmaxxBsecond} term that is always positive.
	Our findings highlight the importance of the Berry curvature on various magnetotransport in magnetic WSMs beyond the anomalous Hall effect.
	
	\vspace{5mm}
	This work was supported by JST FOREST Program Grant Number JPMJFR202N and by JSPS KAKENHI Grant Numbers JP22K18967, JP22H04471, JP21H01804, JP22H04501, JP19K14649, JP23K03275, JP22K20353, JP23K13666 from MEXT, Japan, and also by 2022 Yoshinori Ohsumi Fund for Fundamental Research. A. N. also acknowledges the financial support from Advanced Research Center for Quantum Physics and Nanoscience, Tokyo Institute of Technology.

\newpage
	
\begin{figure*}
	\begin{center}
		\includegraphics*[width=17cm]{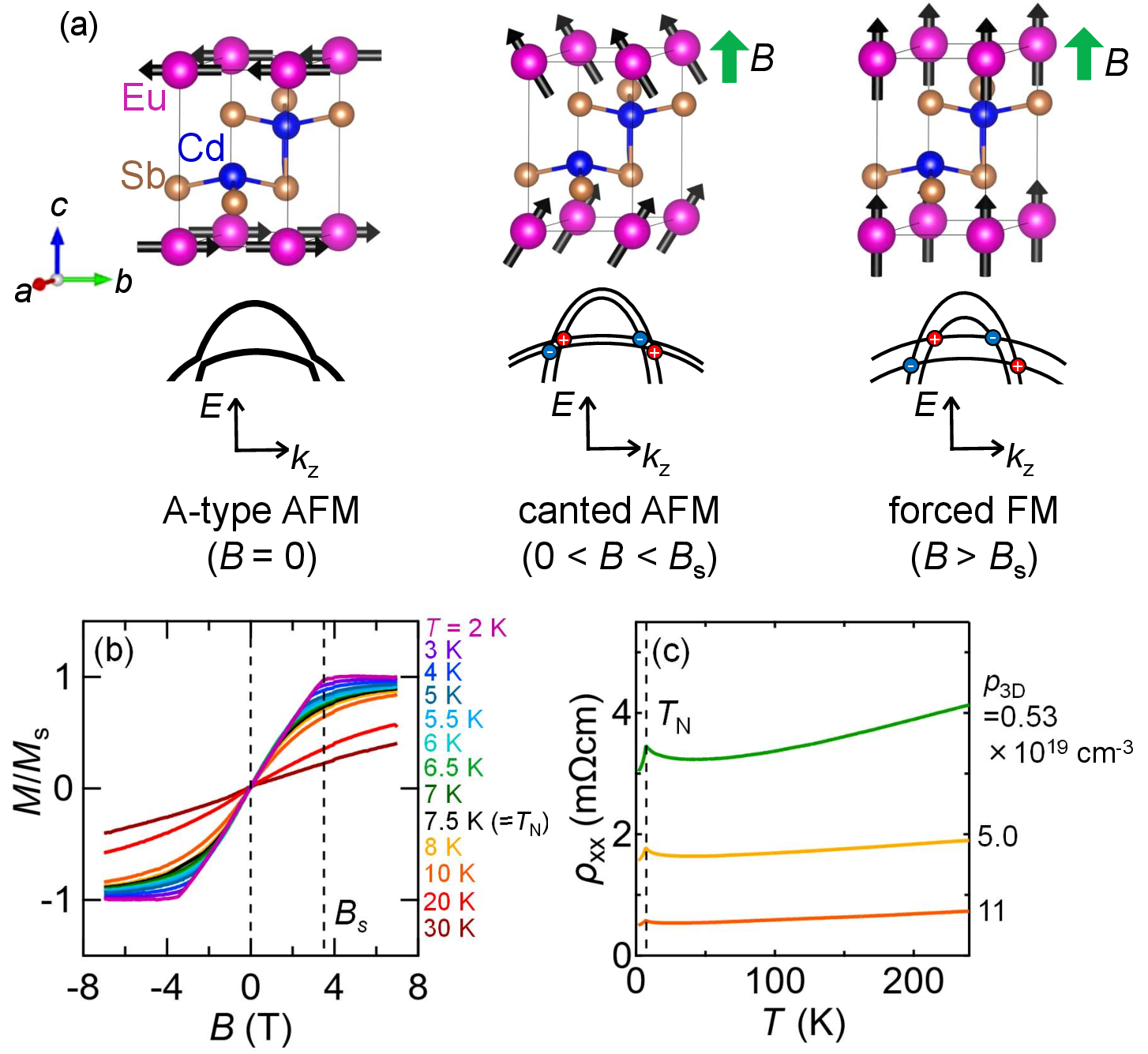}
		\caption{(a) Magnetic orderings and corresponding schematic band structures in {\ECS}, which exhibits a simple magnetization process under the out-of-plane magnetic field, from the in-plane A-type antiferromagnetic (AFM) ordering at zero field to the forced ferromagnetic (FM) ordering above the saturation field $B$$_{\rm{s}}$ \cite{EuCd2Sb2_1, EuCd2Sb2_2}. Upon increasing the field, Weyl points with positive or negative chirality are formed and shifted with the Zeeman band splitting. (b) Magnetization curves measured with sweeping the out-of-plane field at various temperatures in the {\ECS} film with a carrier density of $p$$_{\rm{3D}}$ = \num{0.53e19} \si{cm^{-3}}. (c) Temperature dependence of longitudinal resistivity $\rho_{xx}$, showing a clear kink at the N\'{e}el temperature of {\TN} = 7.5 K.}
		\label{fig1}
	\end{center}
\end{figure*}
	
\begin{figure*}
	\begin{center}
		\includegraphics*[width=18cm]{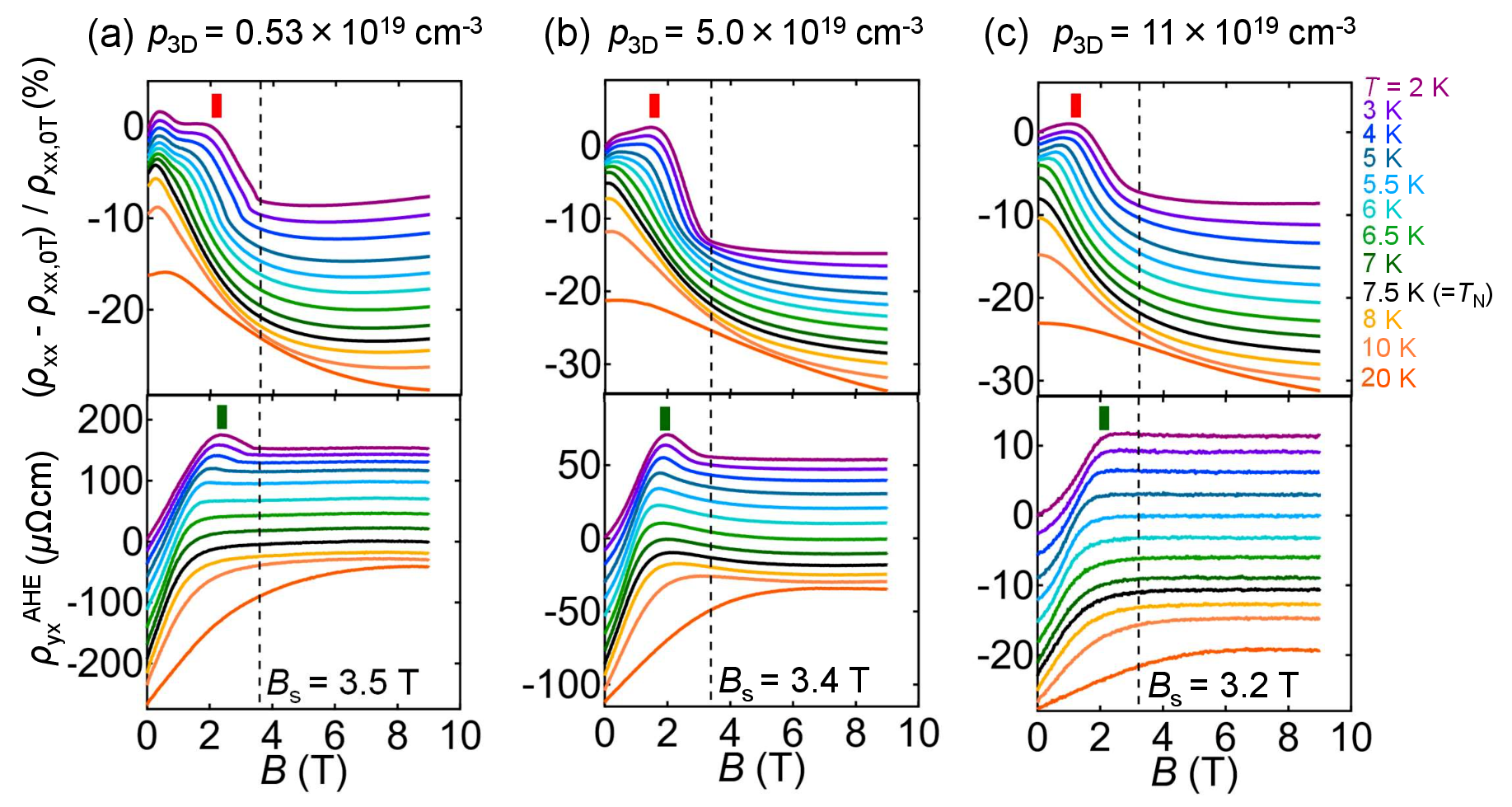}
		\caption{Magnetoresistivity ratio {\magnetoresistivityratio} and anomalous Hall resistivity $\rho_{yx}^{\rm{AHE}}$ measured at various temperatures from 2 K to 20 K in {\ECS} films with carrier densities (a) $p$$_{\rm{3D}}$ = \num{0.53e19} \si{cm^{-3}}, (b) \num{5.0e19} \si{cm^{-3}}, and (c) \num{11e19} \si{cm^{-3}}. A bar highlights the hump position at 2 K. Curves are vertically offset for clarity.}
		\label{fig2}
	\end{center}
\end{figure*}

\begin{figure*}
	\begin{center}
		\includegraphics*[width=14cm]{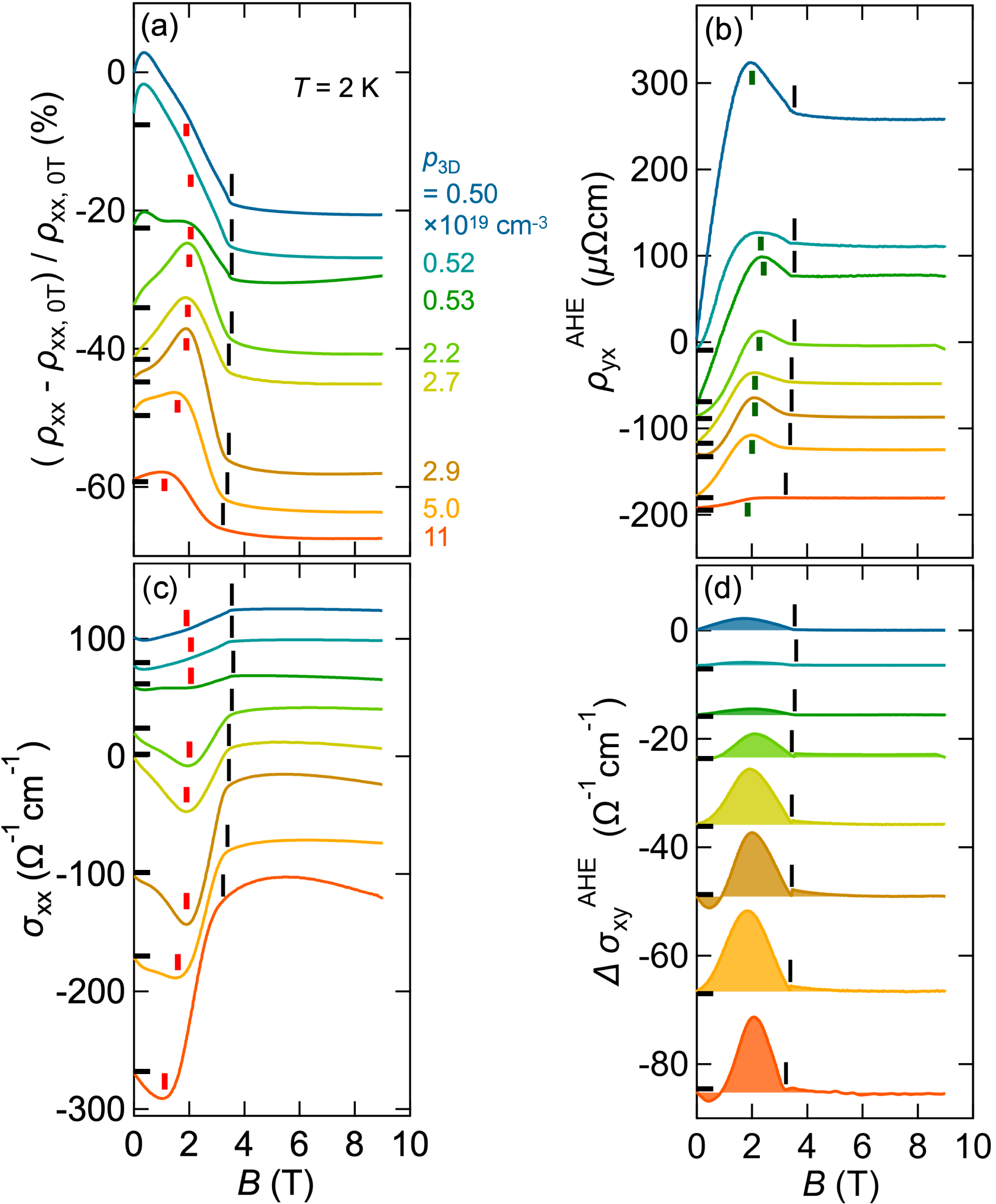}
		\caption{\color{black} (a) Magnetoresistivity ratio {\magnetoresistivityratio} and (b) anomalous Hall resistivity $\rho_{yx}^{\rm{AHE}}$, taken for {\ECS} films with carrier densities ranging from $p$$_{\rm{3D}}$ = \num{0.50e19} \si{cm^{-3}} to \num{11e19} \si{cm^{-3}} at 2 K. (c) Converted magnetoconductivity $\sigma_{xx}$ and (d) nonmonotonic component of the anomalous Hall conductivity {\dsigmaxyAHE} obtained from $\varDelta\rho_{yx}^{\rm{AHE}}$, which is resulted by subtracting the magnetization-proportional component from {\rhoyxAHE}. The saturation field $B_\mathrm{s}$ determined from the magnetoresistance is indicated by a thin vertical bar for each data. Curves are vertically offset for clarity. \color{black}}
		\label{fig3}
	\end{center}
\end{figure*}

\begin{figure}
	\begin{center}
		\includegraphics*[width=12cm]{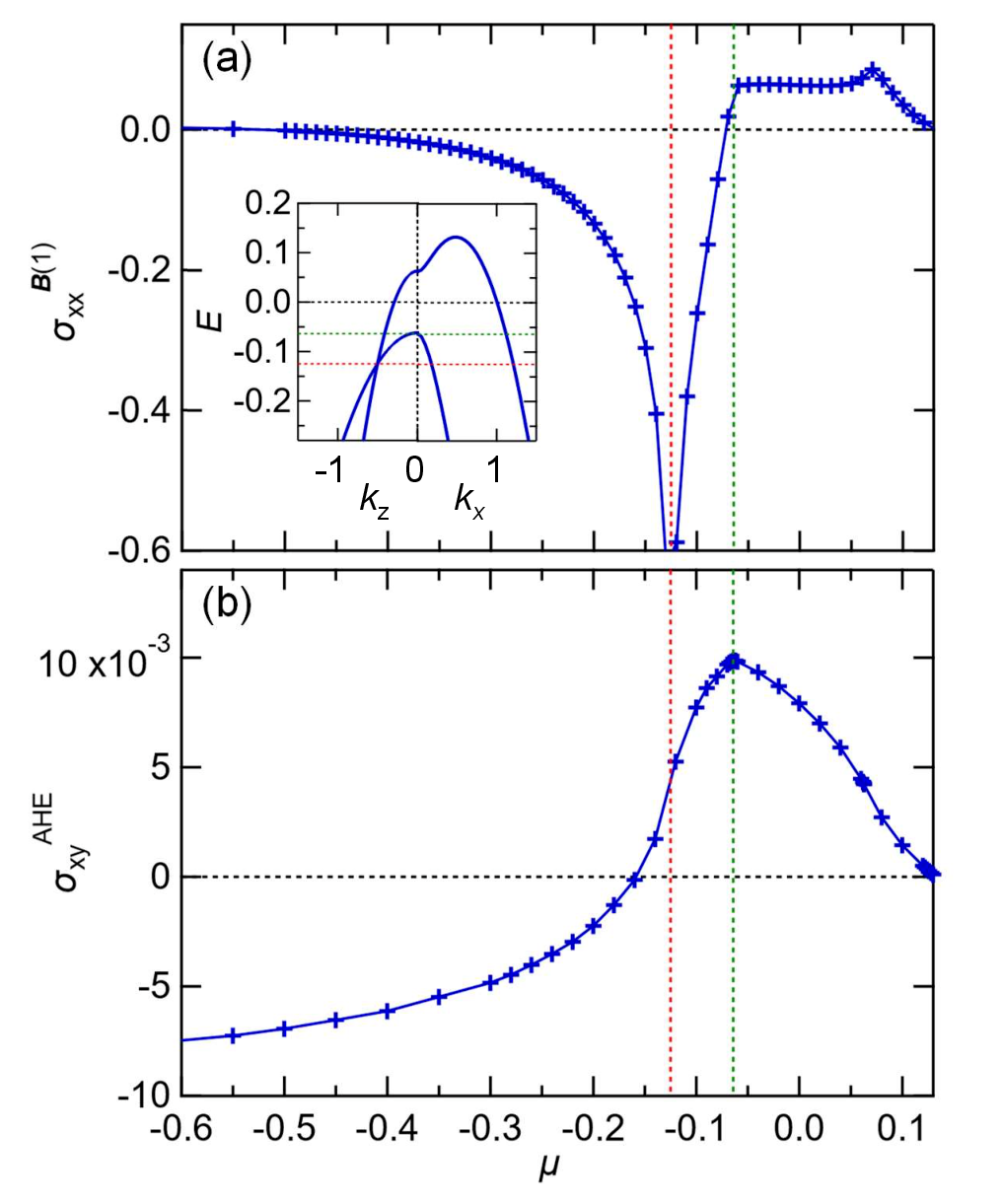}
		\caption{Chemical potential dependence of (a) magnetoconductivity $\sigma_{xx}^{\bm{B}(1)}$ and (b) anomalous Hall conductivity $\sigma_{xy}^{\rm{AHE}}$ calculated for the dipole model. \color{black} The model band structure along the $k_z$ ($k_x=k_y=0$) and $k_x$ ($k_y=k_z=0$) directions is shown in the inset of (a). \color{black} The Weyl points energy and the energy where {\sigmaxyAHE} takes its maximum are indicated by red and green dashed lines, respectively. The results are for $m=1$, $v=1/2$, $k_{0}=1/2$ in Eq. (6).}
		\label{fig4}
	\end{center}
\end{figure}

\end{document}